\documentclass[11pt]{article}
\usepackage{epsf,graphics}

\def\beq{\begin{equation}}
\def\eeq{\end{equation}}
\date{\today}

\begin{document}
\begin{center}
{\large\bf On the phase transition of the 3D random field Ising model}\\[.3in]
  {\bf  Marco Picco$^{a,b}$ and Nicolas Sourlas$^{c}$} \\
  $^{a}$ Sorbonne Universit\'es, UPMC Univ Paris 06, UMR 7589, LPTHE, F-75005, Paris, France\\
  $^{b}$ CNRS, UMR 7589, LPTHE, F-75005, Paris, France \\
  $^{c}$ Laboratoire de Physique Th\'eorique de l'Ecole Normale Sup\'erieure
         \footnote{Unit{\'e} Mixte de Recherche du CNRS et de
         l'Ecole Normale Sup\'erieure, associ\'ee
         \`a l'Universit\'e  Pierre et Marie Curie, PARIS VI.},\\
  24 rue Lhomond, 75231 Paris CEDEX 05, France

\today

\end{center}
\vskip .15in
\centerline{\bf ABSTRACT}

\begin{quotation}

We present new numerical simulations of the random field Ising model in three dimensions at zero temperature. 
The critical exponents are found to agree with previous results. We study the magnetic susceptibility by applying a small 
magnetic field perturbation. We find that the critical amplitude ratio of the magnetic susceptibilities to be very large, equal 
to $233.1 \pm 1.5$. We find strong sample to sample fluctuations which obey finite size scaling. The probability distribution 
of the size of small energy excitations is maximally non-self averaging, obeying a double peak distribution, and 
is finite size scaling invariant. We also study the approach to the thermodynamic limit of the ground state magnetization 
at the phase transition. 

\end{quotation}

Despite numerous efforts, the three dimensional random field Ising model (3DRFIM) is not yet fully understood. 
It is the only case where the perturbative renormalization group (PRG) can be analysed to all orders of 
perturbation theory \cite{Y,PS1} but the PRG fails in the 3DRFIM case.  Several attempts were made to explain this 
discrepancy \cite{PARISIA,mezyoung,mezmon,bredom1,bredom2,PS0,TT}.
PRG predicts dimensional reduction which turns out not to be true. 

Many numerical simulations have been performed to understand the nature of the phase transition. 
The problem has been studied extensively thanks to zero temperature numerical simulations. 
There is a very fast algorithm to find exact ground states \cite{O,AdA,AAS,N}.
In this paper we report new numerical simulations of  the 3DRFIM on cubic lattices with periodic 
boundary conditions and a Gaussian distribution of the random field. 
Many different authors have performed such simulations looking at different aspects of the 
problem.
Until recently the most extensive simulations were those of Angl\`es d'Auriac and Sourlas (AdAS) \cite{AAS}, 
Hartmann and Young (HY) \cite{HY}, Middleton and Fisher (MF) \cite{MF}, and Wu and Machta (WM) \cite{WM}.
When our work was at its final stage we became aware of the work of Fytas and Martin-Mayor (FMM) \cite{FMM}.
These authors performed by far the most extensive simulations up to date with a very careful 
analysis, including next to leading corrections to the critical behaviour. They simulate different 
random field distributions and show universality for the different distributions 
provided one takes proper account of non-leading corrections.
When possible we compare our results to these previous works.

The difficulty of the problem is due to the large sample to sample fluctuations.
We perform a finite size scaling analysis of these fluctuations. This 
analysis shows that these fluctuations are not a finite size effect and that 
they persist into the thermodynamic limit. They are largest for the case of the distribution of the sizes of small energy excitations (see later). 

The model we will consider in this paper is defined through the Hamiltonian 
$$
{\cal H} = - J \sum_{<ij>} \sigma_i \sigma_j - \sum_{i} h_i \sigma_i \; ,
$$
with the spins $\sigma_i = \pm 1$ on a cubic lattice with nearest neighbour interactions 
and $h_i$ independent random magnetic fields obtained from 
a Gaussian distribution with zero mean and variance $H^2$. 
At zero temperature the relevant parameter is the ratio 
of $J^2 / H^2 $. One can vary either $J$ or $H^2 $. In this paper we take $H^2 = 1$ 
and we vary $J$. There is a phase transition at $J=J_c$. $J_c$ is 
known from previous simulations. AdAS found $J_c = 0.441 \pm 0.002$ \cite{AAS} 
while HY obtained $H_c = 1/J_c = 2.28(1)$ and MF got $ H_c =2.27 \pm 0.004 $ \cite{MF}.
The agreement is remarkable because of the very different methods used to determine the critical 
point. The most recent simulations of FMM found $H_c=   2.2721 (2) $, 
in agreement with the previous values \cite{FMM}.
There is similar agreement for the value of the critical exponent $\nu$.

In this paper we assume $J_c = 0.44$, i.e. $H_c = 2.2727...$ and $1/\nu= 0.7$. 
In our simulations, we consider large numbers of samples, typically from 4 million for the smallest sizes to $100 000$ for the largest sizes. 
Each sample corresponds to a realisation of random fields $h_i$. For every sample we computed the ground states for several values of the ferromagnetic coupling 
close to $J_c$ and for small translation invariant magnetic perturbations (see later). 

\section{Magnetization}
In our opinion one of the open questions is the value of the magnetization 
$$ m= {1 \over L^3 } \sum_i \sigma_i $$
at the critical point. It was argued in \cite{AAS} that $m$ is discontinuous at $J=J_c $, 
contrary to what is usually believed, while \cite{MF} argue that $m$ is continuous.

We computed the absolute value of the ground state magnetization $m_c$ at $J=J_c$ for sizes 
$ 11 \le L \le  200 $, and the variation of the ground state magnetization when we change the strength of the 
ferromagnetic coupling. We first report the results for $m_c$ averaged over the samples, $ \overline{m}_c $.  
They are shown in table 1. We see that $\overline{m}_c$ is very large and decreases only slowly  with the lattice size. 
It still remains large even for $L=200$. 

The question is what is its value at the infinite volume limit. We fit the data for $ L_{min} \le L \le 200 $.
It turns out it is not easy to extrapolate to $ L \to \infty $.  The results one gets depend on the hypothesis one makes when extrapolating. 
Different scenarios for approaching the thermodynamic limit can be considered: 
power law behaviour, $\overline{m}_c \sim L^{-{\beta / \nu }} $, logarithmic behaviour, $\overline{m}_c \sim (\log L)^{ - \alpha} $, or 
convergence to a constant $\overline{m}_c \sim const. + L^{ - p }$. 
The results also depend on whether one includes, or not, subdominant corrections,  and on the choice of $ L_{min} $. 
We have tried several scenarios we will detail bellow. 

\begin{table}
\caption{Magnetization $m(L)$ at the critical point for  various sizes.}
\begin{center}
\begin{tabular}{ | l || c | c | c | }
\hline
$ L$ & $m(L) $ & $dm$ \\ 
 \hline
 11  &  0.974303   &  0.000013    \\ 
 16   &  0.968823   &  0.000013   \\ 
 20   &  0.965738  &   0.000014 \\ 
 30  &   0.960296  &    0.000013  \\ 
 42  &   0.955906  &    0.000013 \\ 
 60  &   0.951373  &    0.000026  \\ 
 74   &  0.948683  &    0.000024  \\ 
 90   &  0.946200   &   0.000075   \\ 
 120 &    0.942497  &   0.000054   \\ 
 160  &   0.939018 &   0.000082  \\
 200  &   0.936134 &   0.000085   \\ 
\hline
\end{tabular}
\label{Tablemag}
\end{center}
\end{table}
\begin{enumerate}
\item We assume that 
$$ \overline{m}_c \sim a_0 L^{- \beta / \nu }    $$
i.e. that $ \overline{m}_c$ is zero at the thermodynamic limit and that 
the subdominant corrections are negligible. We obtain a value of $\beta /\nu \simeq 0.013 $ but observe that there are strong 
finite size corrections. This is manifest by monitoring the quality of the fit. We obtain a very large $\chi^2$ per degree of freedom 
if we keep small sizes. It is only after removing data for sizes smaller than $L=30$ that the fit becomes acceptable. 
For $L_{min} \ge 30$ we obtain a value $ \beta / \nu =  0.01335  (40) $ which is stable 
with respect of the choice of $ L_{min} $. Using the same assumptions MF
found $ \beta / \nu =  0.011  (3) $, compatible with us. 
Next we include also subdominant corrections,
\beq
\overline{m}_c \sim a_0 L^{- \beta / \nu } (1+ a_1 L^{-\omega})\; .
\label{Mag1}
\eeq
A fit of the data 
from $L_{min}=11$ up to $L_{max}=200$ gives the values $\beta/\nu = 0.01335 (7)$ 
with $\omega \simeq 1.5 - 2.0$ and a $ \chi^2 < 1.0 $ per degree of freedom. 
Again the value of $ \beta / \nu$ is very stable with respect to the choice of $ L_{min} $. 
Notice that the exponent $\omega$  we obtained is much larger than the one found 
by FMM \cite{FMM}, i.e. $\omega = 0.52 (11)$.

$\beta/\nu$ is extremely small. As MF observe, one would need to go to 
$ 10^{21} < L < 10^{38} $ to see a factor of 2 reduction in the magnetization at the critical 
point! If this were true it would raise the question of whether the thermodynamic 
limit is only of mathematical, and not physical, relevance.

\item Because we found a very small $\beta / \nu $ we tried also to fit under the assumption of 
zero magnetization and a logarithmic approach to the infinite volume limit, including 
subdominant corrections.

\beq
\overline{m}_c \sim {a_0 \over (\log L)^\alpha} (1+ a_1 L^{-\omega})\; .
\label{Mag2}
\eeq
This fit gives $\alpha \simeq  0.1$ and $\omega \simeq 0.5$. The quality of the fit is always very good for any value of $L_{min}$ 
with a small $\chi^2$ per degree of freedom. Note that with this scaling form of the magnetization, the scaling correction exponent 
$\omega$ takes a value very close to the one of FMM \cite{FMM}.  This fact makes this fit attractive to us.
Notice the very small value of the exponent $ \alpha $, i.e. $\overline{m}_c \sim (\log L)^{-0.1} $!!

\item Finally we tested the hypothesis that $ \overline{m}_c$ approaches a value different from zero 
when $ L \to \infty $.
We found a better fit to the data by imposing the obvious condition that $ m \le 1. $ 
 This is done by the change of variables $ \overline{ m } = \tanh(tm) $. We assume
\beq
\label{fit}
tm(L)=tm_0+a_0 L^{-p}+ a_1 L^{-(p+q)} \; .
\eeq

\begin{figure}[t]
\begin{center}
\epsfxsize=450pt\epsfysize=350pt{\epsffile{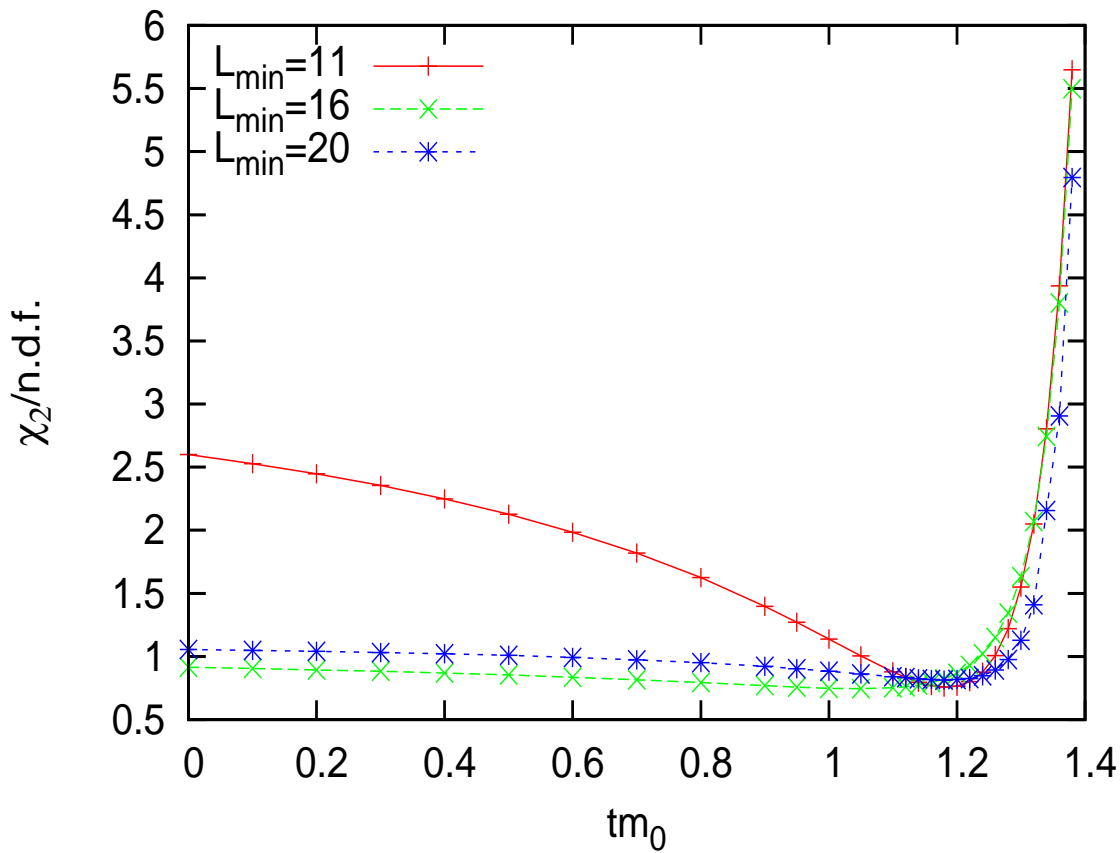}}
\end{center}
\caption{$\chi^2/n.d.f.$ versus $tm_0$ for a fit of $tm(L)$ to the form Eq.(\ref{fit}) with data in the range 
$L=L_{min}-200$ and $L_{min}$ shown in the figure.
}
\label{xima}
\end{figure}

This fit is also excellent from $L_{min} =11 $ to  $L=200 $.
For $L_{min} = 11$ we find $ tm_0 = 1.184 $ or $ m_0 = \tanh(tm_0) = 0.829 $ with $ p = 0.192 $, 
$ q = 1.389 $ and the $\chi^2 $ per degree of freedom equal to $0.884$. If we do not include the small sizes, i.e. $L_{min}=11$, the 
data are overfitted and there is no well define minimum of the $\chi^2$. This is illustrated in figure~\ref{xima}.
From the scaling relation 
$ \overline{m}_c  \sim L^{-\beta /\nu } $, we get that the magnetization critical exponent is
$ \beta = 0 $.

The discontinuity of $ m $ at $J_c$ does not necessary imply a first order transition. 
Because the critical exponent $ \nu $ and the other critical exponents we will measure later in 
this paper take non-trivial values, we argue that the transition is in fact second order with 
$ \beta = 0 $. In the case of a first order transition, we would expect a bimodal distribution 
of the ground state magnetization at the transition point. We found that this not the case, 
in agreement with FMM. They measured the Binder ratio at the transition 
and found $ \overline{m^4} / (\overline{m^2} )^2 = 1.0011 $, i.e. the probability distribution 
of the magnetization is a delta function.

\end{enumerate}

We conclude that if we limit ourselves to the data of table 1 it is not possible to decide with reasonable 
certainty whether the magnetization 
at the critical point $m_0=0$ or not and how the $ L \to \infty $ limit is approached. 
We are inclined to prefer the logarithmic approach  because it gives a value of 
$ \omega $ compatible with FMM.

We next studied  the 
variation of the magnetization $m $ for small changes of the  ferromagnetic 
coupling $ J$ around $J_c$ for different lattice sizes $ L $. 
We will denote in the following  $ dJ= J_c-J $.   For every sample we compute 
ground state magnetizations for different values of $dJ $. (Positive $ dJ$ corresponds to $ J < J_c $.) 
We denote by $ dm $ the difference of the magnetizations, $ dm(dJ,L) = m_c-m(J,L) $ where $ m_c $ is here 
the magnetization of the sample at $ J=J_c$.
We extract the derivative of $ m $ at $J_c $ for size $L$ by 
fitting 
$$    { \overline{  dm(dJ,L) } \over dJ }  = a(L) + b(L) d J + c(L) (d J)^2\; . $$
We define 
$$   \overline{\frac{\partial m (L)} {\partial J }}      = a(L)\; .  $$
We fit the $L $ dependance of $a(L)$ and find $a(L) \simeq L^{1 / \nu }$ with $ 1 / \nu =0.7 $. 
This means that for $J \sim J_c $ and $ J < J_c $ the ground state magnetization 
decreases as $ \simeq L^{1 / \nu }$. 
 
\begin{figure}[t]
\begin{center}
\epsfxsize=240pt\epsfysize=150pt{\epsffile{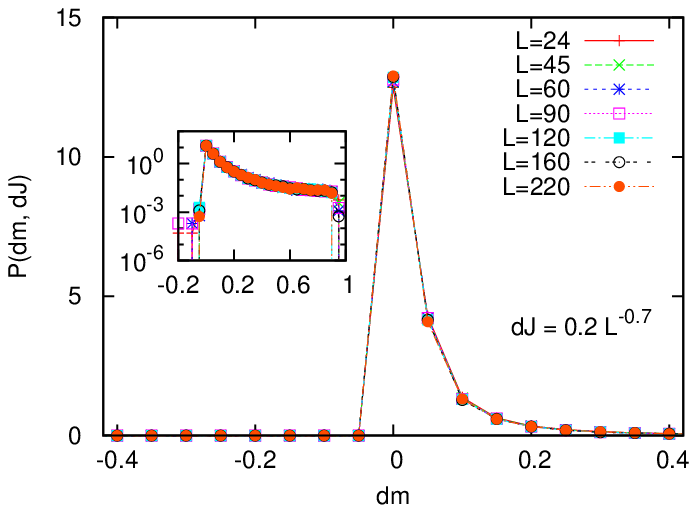}}\epsfxsize=240pt\epsfysize=150pt{\epsffile{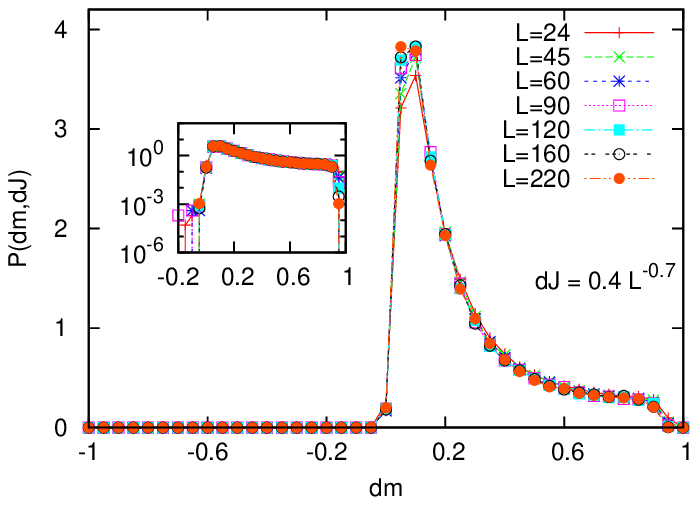}}\\
\epsfxsize=240pt\epsfysize=150pt{\epsffile{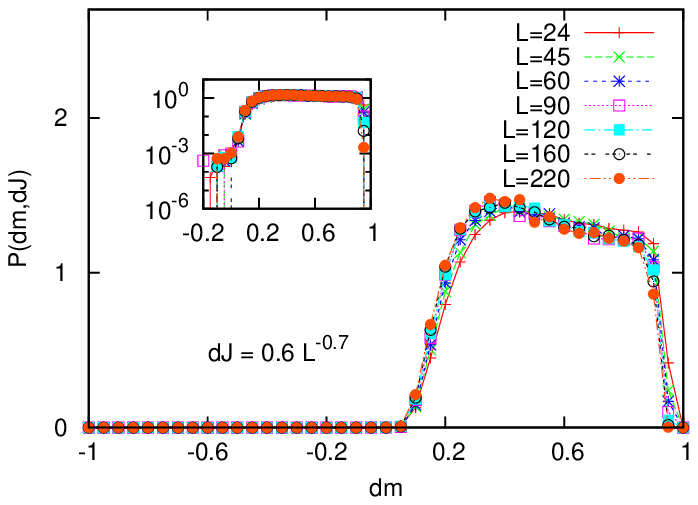}}\epsfxsize=240pt\epsfysize=150pt{\epsffile{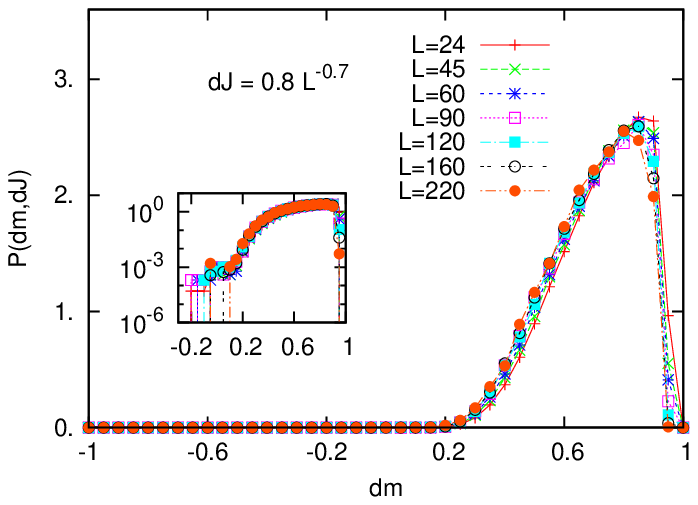}}
\end{center}
\caption{$P(dm, dJ)$ for the values of $dJ$ indicated in the figure. In each case, we show in the inset the same plot with a
log scale on the vertical axis. 
}
\label{xim}
\end{figure}

We show in figure~\ref{xim} the whole probability distribution $ P(dm,dJ,L)$ of $ dm(dJ)$ 
for different values of $dJ $ and sizes $ L $. 
$ P(dm,dJ,L) $ is non-trivial and with an excellent approximation it obeys the finite size scaling relation 
 $$ P(dm,dJ,L) = P(dm,dJ L^{1 / \nu })\; . $$
Not only 
$\overline{ {\partial m }/ {\partial J }}  \sim L^{1 / \nu } $ and becomes infinite in the 
infinite volume limit, but the whole probability 
distribution $ P(dm,dJ,L) $ scales as a function of $ dJ L^{1 / \nu }\;  $.

 We also find a strong violation of self-averaging which is shown 
to persist to  infinite volumes.

The excellent agreement with the finite size scaling 
of $ P(dm,dJ,L) $ presented above and the additional finite size scaling analysis of 
other quantities to be presented below, reinforces our confidence in the value of $J_c $.

\section{Magnetic susceptibility}

To compute the magnetic susceptibility at $J=J_c$, we apply in addition to the random fields a 
very small uniform magnetic field $\delta h $. We compute the corresponding ground state 
magnetizations $\overline{ m_k } $ for different values of $\delta h = \delta h_k $, $k=1,2, \cdots $ and make a fit  
$\overline{ (m_k-m_c) } / \delta h_k = 
\chi(L) + c_1(L)  \delta h_k  + c_2(L)  \delta h_k^2 $. $ m_c$ and $m_k$ are the magnetizations of the same sample 
at $J=J_c$ and $\delta h = 0$ and $\delta h = \delta h_k $ respectively.
This is done for both signs of $\delta h $ and for 
lattice sizes $ L=24,30,60,90,120,160 $. When  $ \delta h_k $ has the same sign as  
$m_c$ we compute the magnetic susceptibility $\chi_+ (L) $. When  $\delta h_k $ has the 
opposite sign of $m_c$ we compute $\chi_- (L)$ and the critical ratio 
$r_c(L)= \overline{ \chi_-(L) / \chi_+(L) } $. 
If the ground state magnetization at the phase transition is large and discontinuous, it 
would have consequences for the behaviour of the magnetic susceptibility at the critical point. 
A small uniform magnetic field $\delta h $ of the same sign as $m_c$ would change very little 
the magnetization, which is already close to one, while a $\delta h $ of opposite sign would have a 
large effect. The critical ratio $r_c(L) $ should be very large.

For $L \ge 24 $ we find that $\overline{\chi_+ (L)} $ and $ \overline{\chi_+ (L) } \sim L^{ \gamma / \nu} $ with 
$ \gamma / \nu = 1.48 \ \pm 0.02 $. 
These errors are purely statistical. The reduced $\chi^2 $ is very small even when including $L = 24 $ and there is 
no need for subdominant corrections. 
For finite temperatures the magnetic susceptibility is 
$ \chi = \beta \sum_{i,j} < \sigma_i \sigma_j > - < \sigma_i > < \sigma_j > $. 
At zero temperature $ < \sigma_i \sigma_j > - < \sigma_i > < \sigma_j > = 0 $ and 
$ \beta $ is infinite and the above relation is not well defined. 
For a Gaussian distribution of the random fields $ h_i $, it can be shown for finite temperatures, 
by integrating by parts over the Gaussian distribution of the random fields,   
that $ \overline{ \chi } \sim \sum_{i,j} \overline{< h_i \sigma_j > } $. 
Assuming the validity of this relation down to zero temperature, FMM found  
 $ \gamma / \nu =1.4847 \pm 0.0010 $, 
in agreement with our value. We find this agreement very remarkable because the methods 
of computing the exponent are so different.

\begin{figure}[t]
\begin{center}
\epsfxsize=450pt\epsfysize=350pt{\epsffile{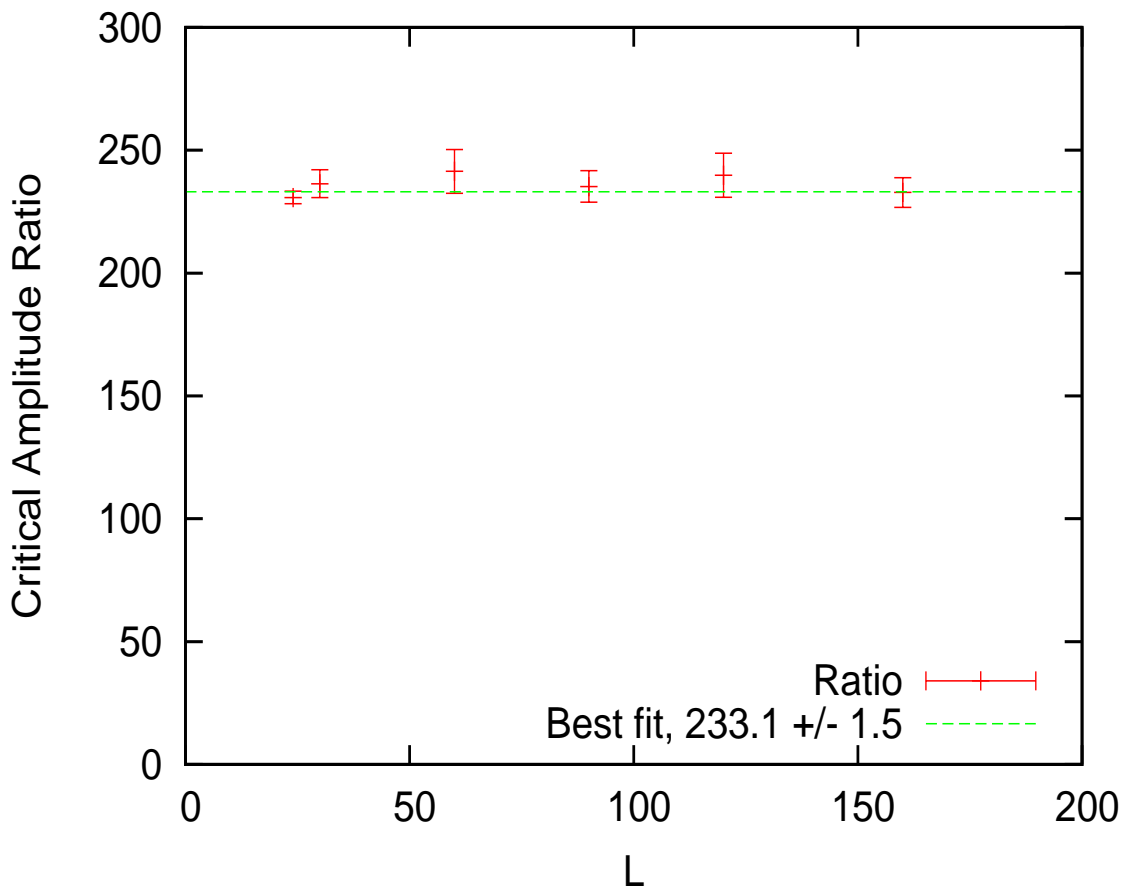}}
\end{center}
\caption{Critical amplitude ratio for the magnetic susceptibility versus the linear size $L$. 
}
\label{Rm}
\end{figure}
Figure~\ref{Rm} shows the critical amplitude ratio $r_c(L)$ for different volumes. 
With a very good approximation $r_c(L) $ is $L$ independent as it should be. 
We find $r_c(L) = 233.1 \pm 1.5 $ which is shown as a dashed line in the figure. 
Notice the very large values of $r_c(L) $ which confirm our expectations. 
We do not know of any other physical system with such large values of critical amplitude ratios. 

\begin{figure}[t]
\begin{center}
\epsfxsize=210pt\epsfysize=150pt{\epsffile{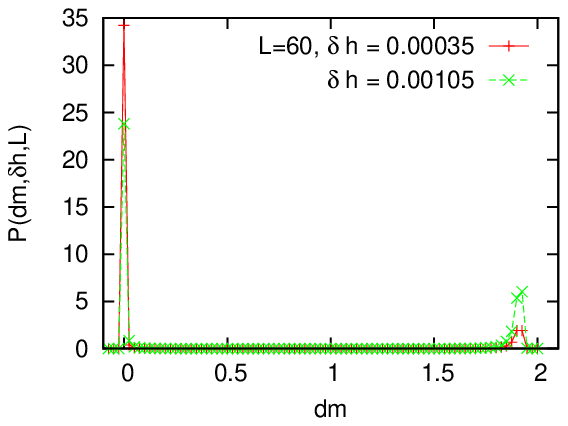}} \epsfxsize=210pt\epsfysize=150pt{\epsffile{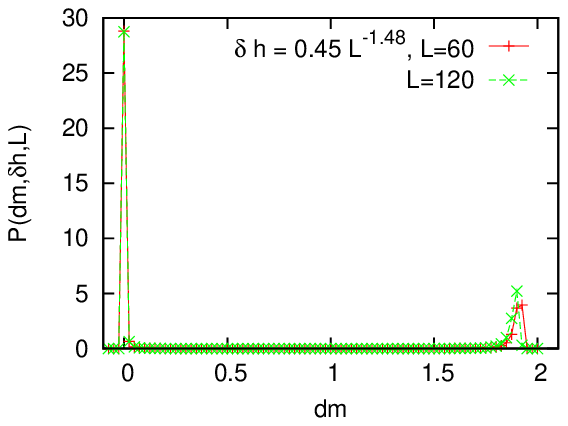}}
\end{center}
\caption{Distribution $P(dm,\delta h,L)$ for $L=60$ on the left and for two values of $\delta h$. On the right, the same 
for $L=60$ and $120$ with  $\delta h = 0.45 L^{-1.48}$.
}
\label{Pdm1}
\end{figure}

We found that $\overline{ (m_k-m_c)} / \delta h_k $ is a very smooth function of $ \delta h_k $, 
and was easy to fit with a polynomial in $ \delta h $.
This hides the fact that $ (m_k-m_c) $ is maximally non self averaging. In fact it has a bimodal distribution!
Only after averaging over samples does it becomes smooth.
We have computed the probability distribution $ P(dm,\delta h,L) $ of $ dm=  m_c-m_k $ for 
different values of $\delta h$ and $L$. The left part of figure~\ref{Pdm1} 
shows $ P(dm,\delta h,L) $ for $ L=60 $ and two different values of $\delta h$ as indicated in the plot.
For both values of $\delta h$, $ P(dm,\delta h,L) $ is a sum of two $\delta$ like functions, 
one centred around $ dm= 0 $, i.e. small change of the ground state magnetization,  and 
the other around $ dm= 2 $, i.e. complete reversal of the system. This feature is true 
for all values of $\delta h$ we studied. Only the height of the two peaks changes with  $\delta h$!
It is remarkable that this happens in such a way that $\overline{ (m_k-m_c)} / \delta h_k $ is smooth and a continuous function of $\delta h$. 

In order to verify the volume dependance  of this violation of self averaging, 
we performed a finite size scaling analysis.
The right part of figure~\ref{Pdm1} 
shows  $ P(dm,\delta h,L ) $ for a fixed combination $ \delta h L^{\gamma/\nu}  $ 
for $ \delta h =0.3 $, 
$ \gamma /\nu=1.48 $  and $ L= 60 $ and $ L= 120 $. 
This shows that the whole probability distribution of sample to sample fluctuations 
obeys finite size scaling, quite a novel result. 
The bimodal nature of  $ P(dm,\delta h,L) $ is not a small size effect but seems to persist 
to infinite volumes. 

Although our estimation of critical exponents is compatible with previous results, 
we think that our study of sample to sample fluctuations adds a new and 
important insight into the nature of the phase transition. 

This study of the magnetic susceptibility also provides information on the low energy 
excited states. It has been proposed \cite{PATA} to add a small repulsion to the ground 
state in order to find low lying excited states. Because $m_c $ is close to one, 
a small $ \delta h $ opposite to  $m_c $ constitutes such a repulsion. 
$ dm $ is a measure of the distance in configuration space between the ground state 
and the first excited states. 
Our calculation of $ P(dm,\delta h,L) $ provides information on their probability distribution.

It is usually assumed that the probability distribution $P(d)$ of the distance $d$ of the first excited state from the ground state 
is narrow and can be described by a single parameter $\overline{d}$, the average over samples of $d$. This assumption 
is important in the phenomenological scaling description of the phase transition in the 3DRFIM. We found that this assumption is not true. 
We think that a more sophisticated analysis, which takes into account the bimodal nature of $ P(d)$, is needed in order to understand, using scaling arguments,
the properties of the phase transition.

\begin{figure}[t]
\begin{center}
\epsfxsize=210pt\epsfysize=150pt{\epsffile{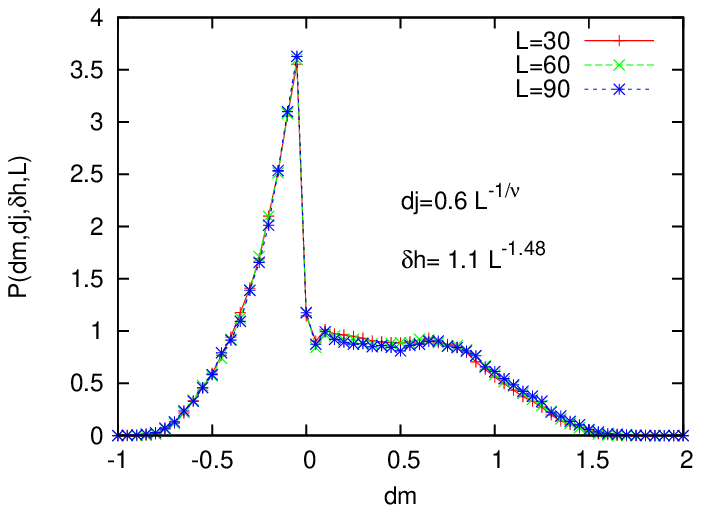}} \epsfxsize=210pt\epsfysize=150pt{\epsffile{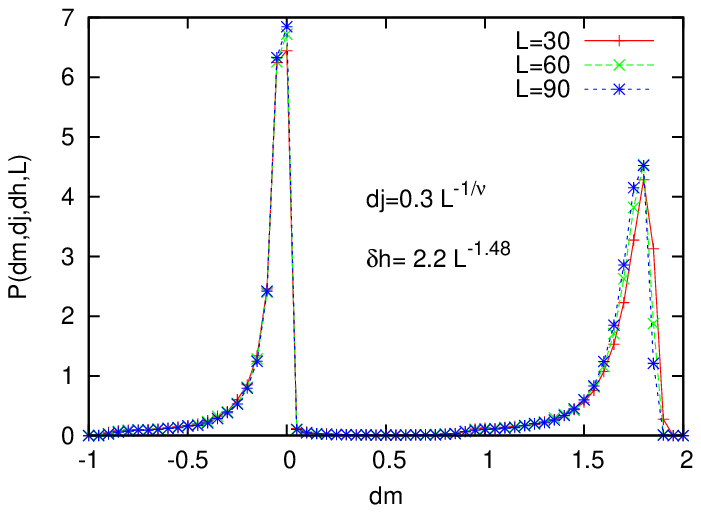}}
\end{center}
\caption{Distribution $P(dm, dJ, \delta h,L)$ for $L=30, 60$ and $90$ for two sets of values of $\delta h$ and $d J$ 
as shown in each figure. 
}
\label{Pdm2}
\end{figure}

We also computed the probability distribution $P(dm,\delta J,\delta h,L)$ of the change 
of the magnetization $dm(dJ,\delta h )=m_c -m(J,\delta h )$, where 
$ m(J,\delta h )$ is the ground state magnetization when, in addition to applying 
a constant magnetic field $ \delta h  $, we change the ferromagnetic 
coupling to $  J= J_c- dJ $. Figure~\ref{Pdm2} shows that there is a double finite size 
scaling invariance, i.e. $P(dm,dJ ,\delta h,L) = 
\tilde{P}(dm,dJ L^{1/\nu},\delta h L^{\gamma/\nu})$. 
In contrast to previous figures, figure~\ref{Pdm2} includes both signs of $\delta h $ (the absolute value of $\delta h$ being the same), parallel and antiparallel 
to the ground state magnetization. 
$P(dm,d J ,\delta h,L) $ remains invariant when we simultaneously change the volume, 
the ferromagnetic coupling and the value of the constant part of the magnetic field 
in such a way to leave both $ dJ L^{1/\nu}$ and $ \delta h L^{\gamma/\nu}$ invariant. 
The scaling function is highly non-trivial and illustrates again the very strong 
violation of self-averaging and its persistance to the infinite volume limit. 
 


\section{Specific heat} 
We also studied the specific heat $ C $ at the phase transition. 
We used a method very similar to the one for the magnetic susceptibility.
For every sample we computed the energy differences $ de_k = e_c -e_k $ 
where $e_c $ is the ground state energy of the sample at the critical point $J =J_c$ 
and $e_k $ the ground state energy at $J=J_k$. We measured $ de_k $ for several values of $J_k$, close to $J_c$.
Let $ dJ_k = J_c-J_k $. For every size $L$ we fitted 
$$ \overline { de_k / dJ_k} = c_0(L) + c_1(L) dJ_k +c_2(l) dJ_k^2 \; ,$$
where $c_0(L)$ is the specific heat at size $L$. 
Our data are very well described by $ c_0(L) = c_0(\infty) + a L^{-p } $, 
For $ 30 \le L \le 120 $, we found 
 $ c_0(\infty) = 2.848 \pm 0.001 $, $p = 0.663 \pm 0.036$ and $ a = 0.179 \pm 0.017 $. 
 The $\chi^2 $ per degree of freedom is $0.06 $. Obviously there is 
 no need for subdominant corrections. 

The specific heat has been discussed at length by Hartmann and Young \cite{HY}.
Although they study smaller systems and their method 
is very different from ours, their result is very similar. 
They found  $ c_0(\infty) = 2.84 \pm 0.05 $, $p = 0.48 \pm 0.03$.
They point out that there are two possible interpretations of this result \cite{HY}.
One consists of interpreting $ c_0(\infty) $ as the regular part 
of the specific heat with the singular part behaving as $ L^{\alpha / \nu } $ 
with $ \alpha / \nu = - 0.663 \pm 0.036  $, where $ \alpha   $ is the 
specific heat exponent. 
The second consists of considering the $ L^{-p }  $
term as a correction to scaling. We remark that, within error bars, $p = 1/ \nu$.
According to this interpretation, the specific heat exponent is $ \alpha = 0.0$. 
 
Assuming our findings, i.e. $ \beta \simeq  0  $ where  $ \beta $ is the magnetization exponent, the value of the 
magnetic susceptibility exponent  $ \gamma  / \nu =1.48 $ and $ \alpha = 0.0 $,  
the scaling relation $ \alpha + 2 \beta + \gamma = 2  $ is satisfied within 
the error bars. For this reason we conclude that $ \alpha = 0.0$. 

\section{Discussion}

Most of our results agree with the previous simulations \cite{HY,MF,FMM}, provided one 
takes into account the relation between critical exponents. 
This is in particular the case for the magnetic susceptibility exponent which is 
related to the correlation function exponents. This agreement 
is very remarkable because our approach is very different from the previous approaches. 
We measure $ \gamma / \nu $ directly, by applying a small magnetic field perturbation, while previous estimations are based on the 
correlation exponent and the relation between exponents. 
The same is true for the specific heat.  

For the first time we present a detailed finite size scaling analysis of the 
sample to sample fluctuations. Absence of self-averaging is not a finite 
size artefact. Our analysis relies on the values of $ J_c $ and $ \nu $. 
We did not estimate $ J_c $ and $ \nu $ ourselves but used the
 values obtained by the previous authors. The validity of finite size scaling of the 
 fluctuations adds further confidence on the values of $ J_c $ and $ \nu $ obtained by these authors.
 
Our  most striking result is probably the bimodal  distribution of the size of low energy excitations.
This was found when, on top of the random fields, we applied 
 very small translation invariant magnetic fields $\delta h $ of sign opposite 
to the ground state magnetization. 
These $\delta h $s play the role of a repulsion to the ground state. 
We found that the probability distribution $P(dm)$ of the resulting 
magnetization difference $dm=m_c - m( \delta h)   $ has a 
double peak shape, one peak around $dm = 0$, and the other around $dm = 2$. 
$dm = 2$ means complete reversal of the spins. 
Because $ m \sim 1 $, $ dm $ measures the distance in configuration space to the ground state.
We found that $P(dm)$ obeys finite size scaling (see figure~\ref{xim}).
Imry-Ma type scaling arguments usually assume a continuous distribution of  the sizes of 
the domains of the excited states. We found that there is no continuous distribution of  the sizes of 
the domains of the excited states. Furthermore their probability distribution, 
which obeys finite size scaling, violates self-averaging in a maximal way. 
We believe that this fact should be taken into account to reformulate the scaling 
picture of the phase transition of the random field Ising model. 

Another very striking result is the very large value of the critical amplitude ratio of the magnetic susceptibility. 

Concerning the ground state magnetization at the phase transition, we found it difficult to reach 
a definite conclusion. We have studied three different scenarios, a) the magnetization 
converging to zero following a power law, i.e. $ \overline {m_c } \sim L^{-\beta / \nu } $, 
b) converging to zero logarithmically, c) approaching a constant when  $ L \to \infty $. 
No one of these scenarios is excluded when one considers only the magnetization data at $J=J_c$,  
not taking into account the other results on the phase transition. In particular the value of the correction to 
scaling exponent $ \omega $ obtained by FMM, strongly favours the logarithmic approach.
On the other hand our results on the magnetic susceptibility critical amplitude ratio seem to favour the 
discontinuous magnetization scenario.

{\bf Acknowledgments}

We would like to thank Victor Martin-Mayor for many stimulating discussions.

\end{document}